\documentclass[reprint,amsmath,amssymb,aps,pra]{revtex4-2}

\usepackage{graphicx}
\usepackage{dcolumn}
\usepackage{bm}
\usepackage{braket}
\usepackage{hyperref}
\usepackage{color}

\begin{document}

\title{Measurement-device agnostic quantum tomography}

\author{Robert St\'{a}rek} 
\email{starek@optics.upol.cz}

\author{Martin Bielak}

\author{Miroslav Je\v{z}ek}

\affiliation{Department of Optics, Faculty of Science, Palack\'{y} University, 17. listopadu 12, 77900 Olomouc, Czechia}


\begin{abstract}
Characterization of quantum states and devices is paramount to quantum science and technology. The characterization consists of individual measurements, which must be precisely known. A mismatch between actual and assumed constituent measurements limits the accuracy of this characterization. We show that such a mismatch introduces reconstruction artifacts in quantum state tomography. We use these artifacts to detect and quantify the mismatch, gaining information about the actual measurement operators. It consequently allows the mitigation of systematic errors in both quantum measurement and state preparation, improving the precision of state control and characterization. The practical utility of our approach is experimentally demonstrated.
\end{abstract}

\maketitle

\section{\label{sec:1}Introduction}

Measurement plays a central role in quantum science and technology~\cite{nielsen_chuang_2010}. However, the implemented measurements often \emph{systematically} differ from the desired ones. Such a discrepancy can be understood as a systematic error. The main focus of this work is on detecting and mitigating these systematic errors. Mitigating systematic error also means gaining information about the measurement operators, which can be used to determine a proper compensation that matches the actual measurements with the desired ones. 
We investigate this problem from the perspective of quantum tomography, one of the most valuable tools in quantum science~\cite{DAriano2003, RehacekParis2004, Teo2015, Eisert2020, Gebhart2023}. It allows the complete characterization of prepared states and implemented quantum circuits, assessing and certifying their quality. Quantum state tomography consists of a suitable set of known measurements on the investigated objects. One can reconstruct the state from the collection of measurement outcomes, the tomogram. The imperfect realization of these measurements causes a mismatch between the assumed and the realized measurement operators. Such a mismatch leads to the wrong interpretation of the acquired tomogram and manifests as a reconstruction artifact. 

Systematic errors in quantum experiments often originate from inaccurate assumptions about the properties or performance of preparation and measurement devices. Such errors may arise from imperfect knowledge of device parameters, deviations of operating conditions from their nominal values, or limitations of prior calibration procedures. For example, the retardance of birefringent wave plates in photonic devices is typically assumed to be exactly $\pi$ (half-wave) or $\pi/2$ (quarter-wave), yet manufacturing tolerances and wavelength mismatches can shift the actual retardance by a few degrees, leading to systematic misrotation of measurement bases. Similarly, in trapped-ion experiments, miscalibrations of the laser intensity or pulse length alter the Rabi frequency, producing consistent over- or under-rotations of qubits. These and related discrepancies between assumed and actual device behavior give rise to systematic errors (bias) that persist across repeated measurements.

The true measurement operators can be fully characterized with \emph{measurement tomography}~\cite{Fiurasek2001, Lundeen2008, Brida2012a, Brida2012b}. While in standard quantum state tomography, we use perfectly known measurements to reconstruct a completely unknown state, in measurement tomography, we use perfectly known states and measure them using an unknown detector, eventually reconstructing the operator that described the measurement. Such a requirement is challenging to meet in experiments, as state preparation can be affected by imperfections and miscalibrations, as well as detection.

Albeit the existence of a scheme for simultaneous reconstruction of a quantum channel with unknown states~\cite{Jezek2003}, the challenge remains. The scheme still requires perfectly known measurements and access to input states before they are affected by the channel. One could model the systematic error in the measurement as a \emph{fixed} unknown channel placed between the state preparation and the detector. However, this approach would be very limiting and would not cover cases where the error varies for each measurement operator used in state tomography.

A \emph{self-consistent tomography} framework has been introduced to loosen the condition of having perfectly known measurements~\cite{Mogilevtsev2012}. The framework suggests the existence of a self-calibrating state class, which, when measured, could reveal initially unknown information about the states and the measurement device. In the self-calibration process, one transfers the prior information about input states into information about the measurement device. In the original work, the authors provide an example of self-calibration using an on-off detector. Although the work provides a conceptual sketch of a generic self-calibration procedure, it was not applied to the system of qubits, i.e., the family of self-calibrating qubit states was not identified.

The problem of quantum state tomography with imprecisely known measurement devices has been addressed~\cite{Keith2018} by utilizing perfect prior knowledge of a few (six) states used to calibrate the detection. Let us note that with such ideal input states, one could theoretically perform quantum measurement tomography~\cite{Fiurasek2001}, calibrate the detector, and then continue with state tomography. Therefore, such an approach does not address the issue of calibrating the measurement device with imperfectly known probe states.

A work \cite{Eisert2023} addressed this challenge by identifying low-rank states as a family of self-calibrating states. The work provides a general and rigorous discussion of self-calibration. In its simplest form, it optimizes the reconstructed density matrices and the parameters describing systematic errors in the measurement to achieve the best fit of the model to the data. The maximum rank of the probe states must be less than an a priori known number. The error model describes the measurement outcome as a linear combination of perfect outcomes and outcomes corresponding to a changed measurement operator in a Pauli basis. The coefficients in the linear combination form an error vector. Another requirement of the method \cite{Eisert2023} is that the error vector has a known and high degree of sparsity. Therefore, the method is effective in cases where a unitary rotation about one of the three axes is dominant.

So far, we have discussed theoretical works. An experimental demonstration of self-calibration~\cite{Braczyk2012} approaches this problem by introducing unknown parameters of the measurement device as another optimization parameter in the likelihood maximization. Then, with a single state tomography, one can recover the unknown parameter, retardance in this case. Since the measurement device parameter is estimated using a single state used to probe it, it is unclear whether there is some ambiguity and how to treat conflicting results about the parameters of a single measurement device calibrated by two different probe states. The work~\cite{Braczyk2012} also addresses the case of two qubits, by adding two unknown parameters into two-qubit tomography. Such an approach, however, would lead to an exponential increase in resources when scaling to more qubits. Another limitation is that the procedure studies only the case of multiplicative errors and does not discuss a more general class of errors.

In this work, we show that an ensemble of single-qubit states of quasi-equal purity distributed quasi-uniformly on the Bloch sphere forms a class of self-calibrating states. This class is particularly convenient in experiments where states are naturally prepared with quasi-equal purity, such as in photonics and trapped ions. Advantageously, the precise control or characterization of such self-calibrating probe states is not required. We present a novel approach to calibrating quantum measurements and state preparation within a single framework, eliminating the need for physical rearrangements of the experiment. To this end, we exploit artifacts in quantum state reconstruction performed on the self-calibrating probe states. 
We also assess how much noise our method can tolerate and still provide an advantage over the calibration-less approach.
In the multi-qubit setting, we can utilize the method in a fully scalable manner to mitigate systematic errors in each local qubit measurement, thereby effectively reducing errors throughout the entire system. The method also applies to quantum device tomography thanks to the state-process duality~\cite{Jamiolkowski1972, Choi1975, nielsen_chuang_2010}, which allows treating quantum processes as states. Our method is not limited to the determination of an initially unknown parameter shared among all measurement operators in the tomographic scheme; it also allows the determination of parameters related to each measurement operator individually.
The presented scheme provides the characterization of quantum states independently of measurement device imperfections. General concepts are introduced using numerical simulations, and we support our findings with a photonic experiment, surpassing the fidelity reported for measurement self-calibration in previous works~\cite{Braczyk2012}. Moreover, the presented method does not require modification of the used reconstruction algorithm - it can be employed as a black-box function in our procedure.

\begin{figure}[b]
	\centering
	\includegraphics[width=1\linewidth]{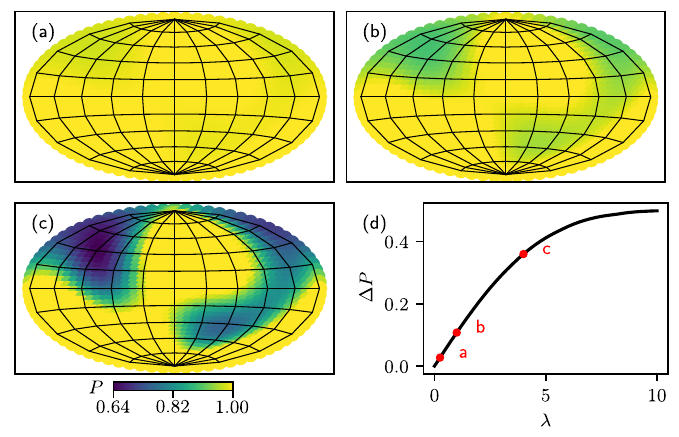}
	\caption{Purity modulation due to the measurement operator mismatch. (a-c) States are represented as points on the Bloch sphere plotted using Hammer projection, and the purity of their reconstructions is color-coded. Although all probe states were prepared with the same purity, the reconstruction using quantum tomography with mismatched measurement operators yields purity modulation. The purity modulation scales with the magnitude of error parameters. Panel (b) shows the purity modulation for randomly selected error parameters described in the main text. Panels (a) and (c) show how the purity modulation, if we scale the error vector by a factor of $\lambda = 0.25$ and $\lambda = 4$, respectively. The black curve in panel (d) depicts the scaling trend, and red dots on the curve mark the  selected values of $\lambda$ plotted in panels (a-c).} 
	\label{fig:1}\
\end{figure}

\section{\label{sec:2}Measurement operator mismatch artifact}
As a first step, let us investigate how the discrepancies between assumed and actual measurement operators manifest in reconstruction artifacts.

In quantum state tomography, many copies of an initially unknown state, described by density operator $\hat{\rho}$, are subject to a series of measurements described by \emph{measurement operators} $\hat{\pi}_{j}$ drawn from a set of positive-operator-valued measure (POVM), which are positive and add up to the identity operator~\cite{nielsen_chuang_2010}. These measurement operators should be tomographically complete, i.e., the density operator must be decomposable into their linear combination~\cite{nielsen_chuang_2010}, and their choice influences the performance of the tomography~\cite{deBurgh2008, Bogdanov2010, Rehacek2015, Martinez2017}. 

The resulting collection of measurement outcomes forms the tomogram. The tomogram and the measurement operators are then input into an algorithm, which outputs the density operator. Let us note that the measurement operators are a necessary input to the reconstruction algorithm, as they provide crucial context for interpreting the tomogram~\cite{Hradil2004}. Here, we use the maximum-likelihood reconstruction algorithm~\cite{Jezek2003, Hradil2004}. If the actual measurement operators $\hat{\pi}_j$ differ from those used in the reconstruction method, $\hat{\pi}'_j$, that is $\hat{\pi}_j \neq \hat{\pi}'_j$, artifacts appear.

To demonstrate the behavior in the core principle of our method, we use pure probe single-qubit states and projective measurements onto the eigenstates of Pauli operators, called Pauli tomography. We must state that this choice is arbitrary and the presented framework is independent of a particular POVM choice. The projective measurement operators are realized by two rotation angles, corresponding to rotation about the $y$- and $z$-axis and subsequent projection onto computational basis state $|0\rangle\langle0|$. Throughout the article, we use the computational basis $\{|0\rangle, |1\rangle\}$. 
The following operators describe the constituent measurements
\begin{equation}
\hat{\pi}_j = \hat{R}_{z}^{\dagger}(\phi_j)\hat{R}_{y}^{\dagger}(\theta_j)|0\rangle\langle 0 | \hat{R}_{y}(\theta_j)\hat{R}_{z}(\phi_j), 
\end{equation}
where $\hat{R}_j(\alpha) = \exp(i\hat{\sigma}_j \alpha/2)$ is the rotation generated by the Pauli operator. Then, the corresponding detection probability is given by the Born rule,
\begin{equation}
p_{j} = \mathrm{Tr}[\hat{\rho}\hat{\pi}_j].
\end{equation}
The angles $\theta_j$ and $\phi_j$ for Pauli tomography are provided in Table~S2 of the Supplemental Material. Assume that due to \emph{systematic} experimental imperfections, the actual angles are 
\begin{eqnarray}
\tilde{\theta}_j = \theta_j + \delta_j,\\
\tilde{\phi}_j = \phi_j + \epsilon_j,
\end{eqnarray}
i.e., we introduce \emph{systematic} additive errors for each measurement operator. We further assume that the parameters $\delta_j$ and $\epsilon_j$ remain effectively static throughout the calibration and evaluation, i.e., they do not exhibit stochastic fluctuations or temporal drift, or if they do, the rate of change is slow enough to be negligible during the calibration and evaluation period. We keep this assumption in the rest of the article.

The error vector $\vec{\delta} = \{\delta_j, \epsilon_j\}$ parameterizes the measurement operators. We numerically simulate tomograms using these true measurement operators for many pure probe states covering the Bloch sphere quasi-uniformly. Then, we reconstruct them assuming the original unperturbed measurement operators $\{\hat{\pi}'_j\}$, i.e., with choice $\vec{\delta}' = \vec{0}$. 
This mismatch causes the state-dependent purity modulation, shown in Figure~\ref{fig:1}(a-c). This example shows the apparent purity modulation for the case of an error vector $\vec{\delta}$, whose elements are drawn from a normal distribution with $\sigma=5$~deg. We quantify the purity modulation in the ensemble of reconstructed states as 
\begin{equation}
\Delta P = \max P - \min P,
\label{eq:purmod}
\end{equation} 
where $P = \mathrm{Tr}\left(\hat{\rho}^2\right)$ is the purity of the reconstructed state.

To show how purity modulation scales with the error vector magnitude, we scale it with a non-negative real factor $\lambda$, i.e., $\vec{\delta} \rightarrow \lambda\vec{\delta}$, and observe the $\Delta P$, as shown in Figure~\ref{fig:1}(d). We see that $\Delta P$ monotonously increases with $\lambda$. Consequently, $\Delta P$ effectively measures how far the assumed measurement operators are from the actual ones. Moreover, the shape of the purity landscape is determined by the \emph{error vector} $\vec{\delta}$.

This effect is not limited to the particular reconstruction method. In Supplementary Material S6, we provide an example of rotating quarter-wave plate polarimetry where we use classical Fourier retrieval of Stokes parameters~\cite{Goldstein} as a reconstruction method. The disagreement between actual and assumed measurement operators causes inconsistency in the tomogram, which results in apparent state-dependent modulation of the degree of polarization, where the degree of polarization can even exceed one.

Apart from the purity modulation itself, other figures of merit, like fidelity, also scale with the error vector. For example, fidelity generally decreases with increasing $\lambda$. We, however, plot only the purity modulation because we intend to use it as the cost function for optimization.

Since $\Delta P$ quantifies the severity of operator mismatch, we now leverage it to identify the true measurement operators and mitigate reconstruction artifacts.

\section{\label{sec:3}Artifact mitigation}
We minimize the artifact severity, i.e., purity modulation~(\ref{eq:purmod}), by varying the measurement operators \emph{used to reconstruct the probe states}. After recording all tomograms of the probe states we used, we optimize our \emph{assumption} $\vec{\delta}'$, which specifies the measurement operators used in the reconstruction algorithm. We use a prime symbol to explicitly distinguish the assumed error vector $\vec{\delta}'$ from the initially unknown actual error vector $\vec{\delta}$. 

Once the \emph{assumed parameters} match the \emph{actual parameters}, that is $\vec{\delta}' \rightarrow \vec{\delta}$, the artifact is minimized. Since the perturbed measurement operators no longer add the identity operator, it is crucial to properly renormalize them in the reconstruction~\cite{Hradil2004}. 

Due to the state dependency, the Bloch sphere must be sampled densely enough to detect the mismatch. The probe states should cover the Bloch sphere quasi-uniformly and possess uniform purity; otherwise, their full control is not required. This feature is advantageous in practice because state preparation is never perfect in the experiment. It distinguishes our approach from the schemes that rely on perfect knowledge of the probe states, like quantum measurement tomography~\cite{Lundeen2008}.

Generally, the optimization landscape contains \emph{multiple local minima}, and therefore global optimization must be performed. Optimizers that start from a single point and use gradient descent may converge to a local minimum, resulting in an incorrect error vector. Note that this issue is not related to whether or not the optimizer uses gradients but rather to whether the optimizer aims to find the \emph{global} minimum.

Here, we use the NOMAD implementation of the MADS algorithm~\cite{Audet2022}. The resulting optimum $\vec{\delta}'$ is not unique because any set of measurement operators $\{\hat{U}\hat{\pi}_j \hat{U}^{\dagger}\}$, where $\hat{U}$ is a fixed unitary operation, also minimizes the artifact. These unitary-equivalent solutions could be interpreted as unitary transformations of probe states. Albeit this ambiguity, such a calibration is already applicable in schemes invariant to local unitaries, such as random measurement \cite{Elben2022} or entanglement certification~\cite{Huber2019, Valencia2020, Guehne2023}, where only relative orientations of $\hat{\pi}_j$ operators are important.
Moreover, the unitary operation can be easily determined by additional measurement of two known and non-orthogonal probe states. 

Since the artifact is state-dependent, the probe sampling strategy is important for minimization. The sampling with too few probe states could lead to incorrect error parameters $\vec{\delta}'_{f}$. We found out that 30 quasi-uniformly distributed probe states provide a reliable determination of error parameters in all tested cases. When we used fewer probe states, we encountered incorrect estimates. The incorrect estimate provides reduced apparent purity modulation in the set of probe states, but when it is used on other states, the purity modulation can be even larger than initially.

To test the method numerically, we first randomly generate ground-truth error parameters $\vec{\delta}$. Each element is normally distributed with zero mean value and standard deviation of 10~deg. Then, we produce tomograms corresponding to pure probe states $\rho_k = |\psi_{k}\rangle\langle \psi_{k}|$. For simplicity, we exclude statistical noise from our simulations for now. Parameters $\vec{\delta}'$ assumed in the reconstruction are optimized to minimize the purity modulation $\Delta P$ in the set of reconstructed probe states $\{ \hat{\rho}_j \}$ for $j=1, \dots, 30$, concluding the calibration step.

To verify the correctness of the optimization result, we use the ground truth $\vec{\delta}$ again to generate tomograms of pure test states $\hat{\xi} = |\xi_{k}\rangle\langle \xi_{k}|$, which quasi-uniformly cover the Bloch sphere in 108 points. We reconstruct these states using the optimized parameters $\vec{\delta}'$ obtained in the calibration step and calculate the purity of the reconstructions. As a further test, we assume knowledge of two non-orthogonal probe states and use their reconstructions to determine corrective unitary operation $\hat{W}$. We apply $\hat{W}$ to all reconstructed test states $\hat{\xi}_k$ to eliminate the unitary-related ambiguity. Then, we check their fidelity with the expected ideal test state, $F_k = \mathrm{Tr}[\sqrt{\sqrt{\hat{\xi}_k} \hat{\rho}_k \sqrt{\hat{\xi}_k}}]$. We characterize the test states ensemble with the purity modulation and their lowest fidelity $\min\limits_{k} F_{k}$ to provide the worst-case estimate.

We repeated the entire numerical simulation 100 times, randomly drawing new ground truth for the error parameters $\delta$ in each run. 
The purpose of this randomization is to test the method performance on various possible combinations of deviations to assess its reliability.
The details about optimizer settings are provided in Supplementary Material~S1. When we compare the optimized results to the reference, i.e., the states reconstructed with the assumption of a null error vector, the average purity modulation decreases from $\Delta P = (8 \substack{+3 \\ -3})) \times 10^{-2}$ to  $\Delta P = (1.0\substack{+0.8 \\ -1.0})\times 10^{-3}$. The uncertainty interval spans from 0.158 to 0.842-quantile, corresponding to two standard deviations, and is used due to the skewed distribution of the results. Mainly, the infidelity decreased from $(3\substack{+1 \\ -2})\times  10^{-2}$ to $(3\substack{+5 \\ -3})\times 10^{-4}$, which represents a two orders of magnitude improvement in accuracy. This improvement is crucial for reaching error levels low enough for quantum error correction schemes.

The lower the $\Delta P$ of the probe state ensemble, the better the results we achieved in the test ensemble. A detailed scatter plot and marginal histogram of these numerical results can be found in Supplementary Material~S2.

The results illustrate the feasibility of the proposed calibration approach and its ability to reliably determine the true measurement operators. It is important to stress that the pure probe states and projective (pure) measurements were chosen here for the simplicity of introducing the measurement-device agnostic framework. However, the method is fully applicable in real cases of generic POVM measurements and nonunity purity, as indicated in Section~\ref{sec:6} using  the experimental demonstration.

To further assess the method robustness, we investigated its performance under two practical noise sources: statistical shot noise and environmental decoherence. In the presence of finite measurement statistics, the method begins to outperform the calibration-less case with $10^3$ shots per measurement setting, and closely approaches the ideal (infinite-shot) performance at around $10^5$ shots. These results can be expected, as the statistical uncertainty scales as $1/\sqrt{N}$, where $N$ is the number of measurement shots. When the statistical error approaches or exceeds the magnitude of the systematic errors, the calibration cannot perform effectively.

To model decoherence, we introduced bit-flip and phase-flip noise channels parameterized by mixing factors $p_x$ and $p_z$, respectively. These parameters range from 0 to 1, where 0 corresponds to no noise and 1 corresponds to complete randomization by the respective channel. We found that the method remains effective at moderate noise levels ($p_x = p_z = 0.01$), and only degrades significantly at stronger noise ($p_x = p_z = 0.1$), where its performance becomes comparable to the uncalibrated approach. See Supplementary Material~S3 for details.


In the multi-qubit setting, quantum state tomography is typically performed using local measurements on each qubit. In this case, our calibration method can be independently applied to each local measurement device. This approach enables the calibration to scale linearly in both measurement resources and computational cost, as each qubit requires only a fixed number of probe states and measurement settings (e.g., 30 states measured along six axes). Moreover, when qubits are independently addressable and correlated noise is negligible, calibration across qubits can be executed in parallel, effectively reducing the total measurement time to that required for a single qubit. The post-processing step, based on our optimization protocol, also parallelizes naturally across qubits.

A more detailed discussion is provided in Supplementary Material~S4, where we demonstrate through numerical simulations that calibrating individual qubits leads to improved reconstructions of $n$-qubit states. We also include a use example in the two-qubit photonic experiment~\cite{Horova2022}.

\section{\label{sec:5}Special case: multiplicative error}
An important subclass of the problem is the case of multiplicative measurement operator error, where
\begin{eqnarray}
\tilde{\theta_j} = (1 + \delta)\theta_j,\\
\tilde{\phi_j} = (1 + \epsilon)\phi_j.
\end{eqnarray} 
This case is relevant when we sequentially switch the unitary operations that precede a fixed projector to perform the tomographic measurements. These multiplicative errors could be interpreted as Bloch sphere under- or over-rotation. This approach to quantum state tomography has been reported in various platforms, including optics~\cite{Resch2005, Shadbolt2011}, color-centers in solids~\cite{Zhang2023}, trapped ions~\cite{Krutyanskiy2023}, neutral atoms~\cite{Noguchi2011}, or superconducting qubits~\cite{Steffen2006}, making this problem subclass highly relevant.

Because parameters $(\delta, \epsilon)$ are shared among all projection operators, the optimum search simplifies from 12 parameters to just two. In Supplementary material~S5, we show that with just eight probe states, the optimization landscape possesses a single minimum, and therefore standard gradient-descent optimization is applicable.
We observe that with the less general error model, one can tailor the probe state distribution to achieve the calibration with a smaller probe state ensemble using fewer computational resources. We further illustrate this on an edge example of the phase over-rotation model in Supplementary Material~S5. 

Another advantage in the case of multiplicative errors is that the global optimum is now unique, and the method unambiguously finds the true set $\{\hat{\pi}_{j}\}$ even when there is an unknown unitary operation between preparation and tomographic measurements. And finally, one can take $\epsilon$ and $\delta$ into account when controlling $\theta_j$ and $\phi_j$ to achieve a more precise control of the projections.

Having mitigated systematic errors in the measurement devices, we next extend the method to calibrated state preparation, solving the proverbial egg-chicken problem of characterizing quantum measurement without precisely known quantum states.

\begin{figure*}[t]
	\centering
	\includegraphics[width=0.999\linewidth]{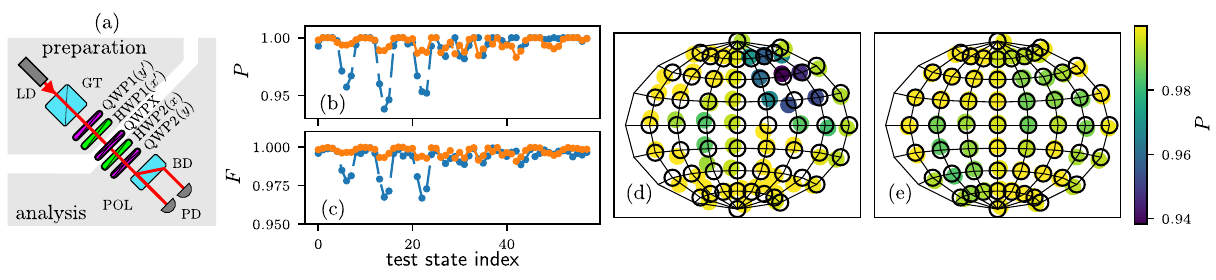}
	\caption{Experimental demonstration for a polarization-encoded photonic qubit. Panel (a) shows the experimental scheme: LD - laser diode, HWP (QWP) - half-(quarter)-wave plate, GT - Glan-Taylor polarizer, BD - calcite beam displacer, PD - photodiode. Panels (b) and (c) show the purity and fidelity of reconstructed test states before (blue) and after (orange) application of measurement-device agnostic calibration. The fidelity of reconstructed states is calculated with respect to theoretical test states. The theoretical test states are depicted as empty black circles on the projected Bloch sphere in panels (d, e). Panel (d) depicts the reference case -- the method was not used, and we assumed perfect waveplates in preparation and tomographic projections. In contrast, panel (e) shows the improved precision of preparation and measurement when we use the method and take into account the retardance deviation. The purity modulation is reduced and the test states match closely the theoretical expectations. In both panels, the reconstructed states are represented on the Bloch sphere using Hammer projection, and their purity is color-coded.}
	\label{fig:2}
\end{figure*}

\section{\label{sec:4}State preparation}
The state preparation typically suffers from imperfections too. Here, we use the symmetry in the Born rule to formally exchange state preparation and measurement to utilize the previously introduced method to correct state preparation. Usually, the qubit is initialized in state $|0\rangle$, then turned into state $|\psi_j\rangle = \hat{U}_j|0\rangle$ by unitary evolution $\hat{U}_j$. For perfect preparation, $\hat{U}_j$ has to be perfectly known. In practice, there might be a slight discrepancy between the desired and the actual evolution. Our calibration reveals this discrepancy and brings information about the actual unitary.

The tomographic projection is also typically realized by some unitary evolution $\hat{V}_k$ and subsequent projection, i.e., $|\pi_k\rangle = \hat{V}_k^{\dagger}|0\rangle$. The state tomogram of the probe state $|\psi_j\rangle$ consists of measured probabilities $|\langle \hat{\pi}_k | \psi_j\rangle|^2 = |\langle 0 | \hat{V}_k \hat{U}_j |0\rangle|^2 $. Here, we assume that the operator $\hat{V}_k$ is precisely known. If not, the calibration procedure described above might be used to improve the knowledge. 

Let us now leverage the Born rule symmetry, i.e. $|\langle \hat{\pi}_k | \psi_j\rangle|^2 = |\langle \hat{\psi}_j | \pi_k\rangle|^2$, to reverse our calibration from measurements to preparations. Physical reversal of the process consists of preparing states $\hat{V}_k |0\rangle$ and projecting them onto states $\hat{U}_j^{\dagger}|0\rangle$, obtaining tomogram elements $p_{kj} \propto |\langle 0 | \hat{U}_j \hat{V}_k |0\rangle|^2$. We can analyze the tomograms to improve our knowledge of the measurement operator, i.e., $\hat{U}_{j}^{\dagger}|0\rangle\langle 0| \hat{U}_j$, related to state preparation in the non-reversed situation. 

The reversal can be realized in the experiment without its physical rearrangement, by performing projections onto the original probe states in the tomographic device and preparing states corresponding to the original tomographical projectors instead of the original probes. Either way, we gain information about $\hat{U}_j$, increasing the accuracy of state preparation.

As an example, first recall the situation from Section~IV. To calibrate six measurement operators, we applied them to eight probe states. These measurements were implemented by unitary rotations $V_k$, followed by projection onto the state $|0\rangle$. The calibration effectively revealed information about the $V_k$, i.e. parameters $\delta$ and $\epsilon$ which can be used to more precisely control the projections.

Next, one would prepare six new probe states corresponding to the original projection $V_k|0\rangle$. These states carry information about systematic errors in the preparation. Each of these would be projected onto the original eight probe states $ U_j|0\rangle$. The resulting tomograms contain readings corresponding to imprecise preparations measured with precise measurements. It could then be analyzed to gain information about the parameters of the preparation device. This information can be used to correct the state preparation, as we will demonstrate experimentally in the following section.

\section{\label{sec:6}Experiment: polarization state tomography}
We experimentally demonstrate the method in the waveplate-based polarization state tomography scenario. At the end of the process, we will improve our knowledge of the waveplate retardance and consequently mitigate the reconstruction artifact. 

The experimental setup is depicted in Figure~\ref{fig:2}(a). We used a continuous-wave 810~nm thermally-stabilized fiber-coupled laser diode (LD). The laser beam was decoupled into free space using an 11-mm aspheric lens and horizontally polarized using a Glan-Taylor prism (GT). The waveplates HWP1 and QWP1 effectively control the prepared polarization state. The following operator describes the action of a waveplate:
\begin{equation}
	\hat{W}_{\alpha, \gamma} = |\alpha\rangle\langle\alpha| + e^{-i\Gamma}|\alpha_\bot\rangle\langle\alpha_\bot|,
\end{equation}
where $\alpha$ and $\Gamma$ are its angular position and retardance, and $|\alpha\rangle = \cos{\alpha}|0\rangle + \sin{\alpha}|1\rangle$ is a linearly polarized state, and $\alpha_\bot = \alpha + \pi/2$. To prepare a state located at colatitude $\theta$ and longitude $\phi$ on the Bloch sphere, we rotate the half- and quarter-wave plate to angles $x'$ and $y'$ given by the relations
\begin{eqnarray}
	y' = -\frac{1}{2}\arcsin\left(\sin \theta \sin \phi \right),\\
	x' = \frac{1}{4}\arctan\left(\tan\theta \cos\phi \right) + \frac{1}{2}y + c(\theta),
	\label{eq:hwpqwpprep}
\end{eqnarray}
where $c(\theta) = \pi/4$ if $\theta > \pi/2$ and $c(\theta) = 0$ otherwise. This relation holds for wave plates that have perfect retardance. The deviation from the ideal retardance will introduce errors in the preparation. We assumed the perfect retardance in the state preparation stage as our method does not require exact knowledge of probe states. 

To demonstrate the tolerance of the method to a unitary operation between the preparation and tomographic measurement, we add a quarter-wave plate (QWPX) with its fast axis oriented at {-12.5~deg} relative to the horizontal direction. Because this unitary operation can be in principle unknown, we restrict ourselves from using any information about QWPX in the calibration.

The states were analyzed using half-waveplate and quarter-wave (HWP2, QWP2) and subsequent projection onto horizontal and vertical polarization using a calcite beam-displacer (BD). The waveplate settings for the Pauli tomography are listed in Tab.~S3 of the Supplementary Material. All waveplates were mounted into a precision motorized rotation stage that provided bidirectional repeatability of $\pm30$~mdeg. 

We used reverse-biased PIN photodiodes (PD) to measure optical intensity at both outputs of the beam-displacer. We normalized the readings using the sum of the two signals to eliminate the influence of laser power fluctuation.

The action of the waveplate could be seen as the rotation of the Bloch sphere around axis $(\sin(2\alpha), 0, \cos(2\alpha))$ with rotation angle $\Gamma$, where $\alpha$ determines the angular position of the waveplate and $\Gamma$ its retardance. The retardance usually deviates from its nominal value. These deviations manifest as under- or over-rotation of the Bloch sphere. So, in the context of the previous sections, the deviation of the effective wave plate retardance from its nominal value is the origin of the mismatch between the measurement operators. The retardance mismatch can be attributed to the manufacturing tolerances or to the use of a light at a slightly different wavelength than the nominal design value of the wave plate. The exact values of these deviations are initially unknown to us, and we will use the presented method to estimate their values more precisely. We parameterize the measurement operator by precisely controlled waveplate angular positions $x$, $y$, and half- and quarter-wave plate retardance deviations $\delta'$ and $\epsilon'$, respectively.

We sequentially prepare eight distinct polarization states and perform the Pauli tomography for each. As we introduced in previous sections, these \emph{probe states} have similar purity and are quasi-uniformly distributed over the Bloch sphere. We use them to probe the measurement device. Initially, we assume zero deviations, $\delta' = \epsilon' = 0$, and observe the purity modulation $\Delta P = 0.0572(3)$. The number in parentheses is one standard deviation at the last significant digit, e.g. $0.0572(3) = (572\pm 3)\times 10^{-4}$. The uncertainties were estimated using bootstrapping with the assumption of normally distributed photo-current with 0.1\% relative standard deviation, which is the most dominant term and is related to the ammeter precision. The shot noise is negligible at the power level of our laser, compared to the technical noise. Then, we vary parameters $\delta'$ and $\epsilon'$ and update the measurement operators employed in the reconstruction to minimize the purity modulation. The optimal $\delta$ and $\epsilon$ are close to the true values because the reconstruction artifact is significantly reduced to $\Delta P = 0.0140(2)$. We found the optimal parameters to be $\delta' = 4.5(2)$~deg, $\epsilon' = -1.30(4)$~deg. 

We then reverse the process to determine retardance deviations $\tilde{\delta}, \tilde{\epsilon}$ of the waveplates used in the preparation stage. We prepare the six eigenstates of Pauli operators, project each onto eight states corresponding to the original probe state set, and treat this data as before. We minimize the apparent purity modulation from 0.0700(3) to 0.0144(2) to find the optimal parameters $\tilde{\delta} = 4.53(6)$~deg and $\tilde{\epsilon} = -3.58(3)$~deg.

The calibration step revealed the retardance deviation of the wave plates used for preparation and projection. To reconstruct the quantum states, we could directly use this knowledge to update our measurement operators. However, we take one extra step, which allows us to prepare the states with greater precision and control the measurement operators precisely. We took these deviations into account in the verification step and calculated new angular positions of wave plates for tomographic projections and state preparation. Since Eq.~\ref{eq:hwpqwpprep} does not hold anymore, we used numerical optimization to find the desired wave plate settings ($x, y$):
\begin{equation}
	x,\,y = \mathrm{argmax}_{x,y}|\langle \psi| \hat{W}(x, \pi+\delta) \hat{W}(y, \pi/2+\epsilon)|0\rangle|^2.
	\label{eq:wpminprep}
\end{equation}
Similarly, to find the settings to project a state onto state $|\pi\rangle$ we perform this maximization
\begin{equation}
	x,\,y = \mathrm{argmax}_{x,y}|\langle \langle 0| \hat{W}(x, \pi/2+\epsilon) \hat{W}(y, \pi+\delta)|\pi\rangle|^2.
	\label{eq:wpminproj}
\end{equation}

At this point, we have information about actual measurement operators, and we have also calibrated the state preparation. This calibration was done in the presence of a unitary operation between state preparation and measurement, which was introduced by QWPX. This operation was effectively unknown because we did not use any prior information about QWPX during the process.

Now, we proceed to certify the correctness of such a calibration. The calibration should remain valid even if we change the central unitary operation introduced by QWPX. To test this, we removed the QWPX and then prepared an ensemble of 58 test states, depicted as empty circles in Figure~\ref{fig:2}(c,d). 

As a reference, we first assumed perfect waveplates and performed state tomography for each test state. We observed purity modulation $\Delta P = 0.0607(2)$ with purity minimum $P_{\min} = 0.9383(2)$. This modulation indicates reconstruction artifacts. The minimum fidelity of the reconstructed state to the target states was $F_{\min} = 0.96688(4)$. These fidelity values indicate imprecise preparation. 

In contrast, when we used our knowledge of $\delta$ and $\epsilon$ in the reconstruction and preparation, the minimum purity increased to $0.9830(3)$ and the minimum fidelity to $F_{\min} = 0.9907(2)$. The information gained in the calibration clearly improved the reconstruction. The overall fidelity $F = 0.997(3)$ surpasses the value 0.99(1) reported previously by~\cite{Braczyk2012}. Note that this fidelity is calculated relatively to the expected ideal state and was evaluated on a set of 58 test states, independently of calibration. This is in contrast to the previous demonstration, where only 14 states were used, and these states were also used for calibration.

The results in Figure~\ref{fig:2} indicate improved quality of state preparation and tomographic measurements when we use the knowledge of $\delta$ and $\epsilon$ to improve both preparation and measurement. The small remaining artifacts are present due to the limited precision of the learned retardance deviations and also the repeatability of the wave plate rotation. 

Having demonstrated experimental feasibility with polarization-encoded optical qubits, we now turn to on-chip photonic qubits and illustrate how our calibration method applies to path-encoded qubits in this increasingly relevant setting.

\section{\label{sec:s7}Path-encoded qubits on a photonic chip}

\begin{figure*}[t]
	\centering
	\includegraphics[scale=1]{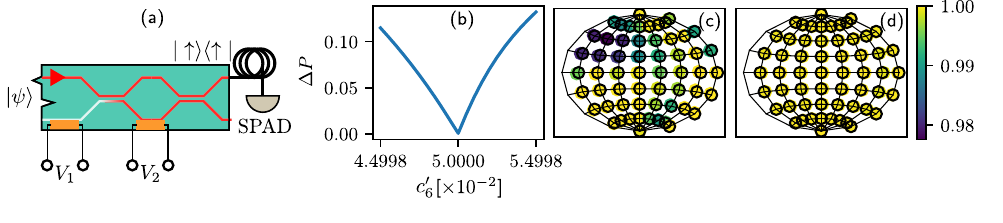}
	\caption{Application of the method for an on-chip photonic path qubit.
		(a) Projection apparatus for a path qubit. Applied voltage $V_{1,2}$ on heater elements (orange) causes thermal-based phase shift in the waveguide interferometers. These shifts determine the measurement projector. The photons in the upper output path are detected with a single-photon detector (SPAD). 
		(b) 1D-section of 6D function $\Delta P(c'_1, \dots, c'_6)$. The value of $\Delta P$ at the initial assumption  $c'_j$ is $1.8 \times 10^{-2}$ on the ensemble of probe states. Minimum $\Delta P = 4.5 \times 10^{-4}$ lies at the true value $c_6$.
		(c) Initial mismatch results in a state-dependent purity artifact. We plot test states on the Bloch sphere as circles and color-code their reconstructed purity.
		(d) When we consider the true retardance, the artifact vanishes.
		}
	\label{sfig:chip}
\end{figure*}

As another example, we numerically illustrate the measurement-device agnostic calibration for tomography of photonic on-chip path qubits~\cite{Shadbolt2011, Carolan2015, Flamini2015, Renema2023}. The qubit is encoded into the photon propagation in one or another waveguide (states $\ket{\uparrow}$ or $\ket{\downarrow}$), as depicted in Figure~\ref{sfig:chip}(a). The local heating of one waveguide controls the phase between the two paths. The heating is typically achieved by an electrical current flowing through a small heater element. A Mach-Zehnder interferometer serves as a variable coupler, and its interferometric phase controls the coupling ratio. Detection of photons in a single optical path implements a projective  measurement of the path qubit. The circuit in Figure~\ref{sfig:chip}(a) implements a projection onto state $\ket{\pi}$, whose colatitude and longitude on the Bloch sphere are controlled by optical phases. The relation between optical phases $\varphi_{1,2}$ and voltages $V_{1,2}$ applied on the heating elements is approximated with polynomials~\cite{Carolan2015}
\begin{eqnarray}
	\label{eq:pathphase}
	\varphi_1(V_1) = c_1 + c_2 V_1 + c_3 V_1^2, \\
	\varphi_2(V_2) = c_4 + c_5 V_2 + c_6 V_2^2. \\
\end{eqnarray}

Here, the voltages $V_{1,2}$ are controlled perfectly, but parameters $c_{j}$ are known only approximately. The other imperfections are neglected. We have increased the number of probe states to 22 to compensate for the higher number of optimization parameters. The simulated experiment, depicted in Fig.~\ref{sfig:chip}(c), shows the measurement operator mismatch artifact. We search the six-dimensional parametric space to find the point of minimal $\Delta P(c_j ')$ where the systematic error is mitigated. Panel (b) shows a one-dimensional section illustrating that the minimal $\Delta P$ corresponds to the true value $c_6$. Taking the true values $c_{j}$ into account mitigates the reconstruction artifact, as shown in panel (d).

\section{\label{sec:7}Conclusion}
The mismatch between actual measurement operators and their theoretical counterparts assumed in reconstructing quantum states or devices introduces reconstruction artifacts. We showed how to leverage these artifacts to reveal the actual measurement operators and mitigate systematic errors in the tomography and state preparation. We experimentally applied this method to Pauli tomography of polarization-encoded photonic qubits and numerically analyzed several other measurement schemes. The proposed method makes the tomography independent of the measurement device. The main advantage is that the method does not require perfect knowledge or control of the probe states, which is virtually impossible to obtain. The method can calibrate already existing tomographic apparatus without the need for individual characterization of its constituent components. This is particularly useful in the case where the experimental setup is monolithic and individual components cannot be calibrated individually, e.g. in the case of integrated circuits. One can also calibrate the state preparation using the same method.
Furthermore, the presented approach applies locally to individual parties of a larger multiparty system; i.e. it is fully scalable.

In summary, we developed the framework for accurate and scalable calibration of quantum measurement devices and also quantum state preparation, which is critical in any applications of quantum science and technology.

The application of the reported measurement-device agnostic quantum measurement goes beyond the full quantum tomography~\cite{Jezek2003, Hradil2004}. The same method will find its use for mitigation of measurement errors in various approximative and more scalable approaches, such as Monte-Carlo sampling \cite{Flammia2011, Micuda2013, Starek2016, Micuda2015}, permutationally invariant tomography \cite{Toth2010}, compressed sensing \cite{Eisert2010, Blatt2017}, etc. The method can also improve the accuracy of building and characterization of the photonic and quantum circuits in complex media \cite{Gigan2019, Malik2024}. Finally, the reported approach also applies to experimental realizations of entanglement certification and quantification \cite{Huber2019, Valencia2020, Elben2022, Koutny2023, Guehne2023}.

The supporting data for this article are available from the Zenodo repository~\cite{zenodo}.

\begin{acknowledgments}
We acknowledge the support of the Czech Science Foundation under grant No.~21-18545S. M.~B. acknowledges the support of Palacký University under grants No.~IGA-PrF-2024-008 and 
IGA-PrF-2025-010. We acknowledge the use of cluster computing resources provided by the Department of Optics, Palack\'{y} University Olomouc. We thank J. Provazn\'{i}k for maintaining the cluster and providing support. 
\end{acknowledgments}

\bibliography{reference.bib}

\end{document}


\title{Measurement-device agnostic quantum tomography -- Supplementary material}

\author{Robert St\'{a}rek} 
\email{starek@optics.upol.cz}
\author{Martin Bielak}%
\author{Miroslav Je\v{z}ek}
\affiliation{Department of Optics, Faculty of Science, Palack\'{y} University, 17. listopadu 12, 77900 Olomouc, Czechia}

\maketitle

\section{\label{sec:s1n}NOMAD setting in the numerical test}

When we numerically investigated the case of general additive errors, we set the NOMAD~\cite{Audet2022} parameters as listed in Tab.~\ref{stab:1}. Please refer to the NOMAD 4 manual~\cite{nomadmanual} for the meaning of the parameters.  The optimized parameters were bounded to the interval (-29, 29) deg, i.e., roughly three standard deviations of the true parameter distribution. The seed was pseudo-randomly chosen at the time of optimization. The maximum size of the evaluation block was set as twice the problem dimension to allow optimal parallelization of the initial Latin hypercube search. We used the \emph{nomadlad}~\cite{nomadlad} package to use NOMAD from Python. The necessary configuration was to set NOMAD to use multiple initial optimization points instead of just one by not specifying parameter X0 and setting the Latin hypercube search parameter (LH SEARCH) sufficiently high. 

\begin{table}[t]
	\begin{tabular}{|l|l|}
		\hline	
		parameter & value\\
		\hline
		BB OUTPUT TYPE & OBJ \\
		DIMENSION & 12 \\
		BB MAX BLOCK SIZE & 24\\
		MAX BB EVAL & 100 000\\
		SEED & random integer\\
		DIRECTION TYPE & ORTHO 2N\\
		LH SEARCH & 1100 1100\\
		LOWER BOUND & (-0.5, \dots, -0.5)\\
		UPPER BOUND & (0.5, \dots, 0.5)\\
		\hline	
	\end{tabular}
	\label{stab:1}
	\caption{NOMAD (v4.1.0) settings used for the global optimization. Some parameters are not listed because they are not critical for this application.}
\end{table}

\section{\label{sec:svis}Visualization of results from Section III.}
\begin{figure*}[ht]
	\includegraphics[scale=1]{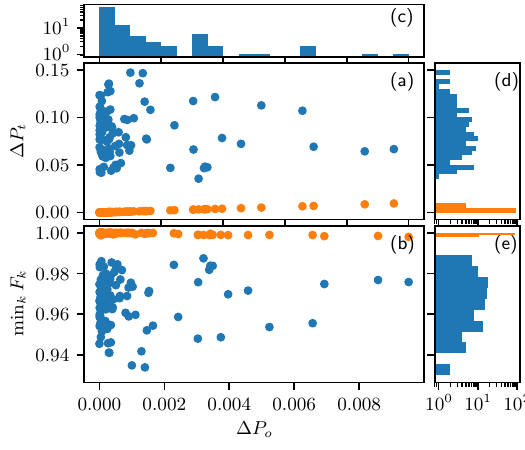}
	\caption{Visualization of results from Section~III of the main text. The figure shows the purity modulation amplitude $\Delta P_{t}$ of 108 reconstructed \emph{test states} (a) and its minimal fidelity (b) versus the purity modulation $\Delta P_{o}$ achieved in optimization over 30 \emph{probe states}. Panels (c), (d), and (e) show the respective histograms in log scale. The blue color in panels (a,b,d,e) represents the reference case, where we assumed $\vec{\delta} = 0$, while the orange color represents the optimized case. Each point represents a single test for a specific randomly chosen ground truth. The smaller $\Delta P_0$ the optimizer achieved, the better performance we observed.}
	\label{sfig:simvis}
\end{figure*}

In Section III of the main text, we have summarized the numerical test using descriptive statistics. Here, in Fig.~\ref{sfig:simvis}, we supplement this information with a scatter plot of the individual realizations and the respective marginal histograms.

\section{\label{sec:noises}Noise analysis}

So far, the performance of the presented method has been evaluated in the absence of noise, i.e., perfect qubits, and in the limit of an infinite measurement shots. It is natural to ask how the method performs under realistic noisy conditions. In this section, we first analyze the influence of shot noise, followed by the impact of environmental noise.

Intuitively, the shot noise introduces random fluctuations in the reconstructed purity. As expected, the smaller these fluctuations, the more accurately we can estimate the true error parameters. To illustrate this behavior, we repeated the simulation described in Section~III of the main text with the following modification: we assume that constituent measurements are performed sequentially and with a finite number of shots. Each element of the resulting tomogram is drawn from a Poissonian distribution. The simulation was repeated for shot counts ranging from $10^2$ to $10^5$ in logarithmic steps. As a reference, we also rerun the simulation with zero noise. For each setting, we took 50 Monte Carlo samples.

\begin{figure*}[ht]
	\includegraphics[width=0.8\linewidth]{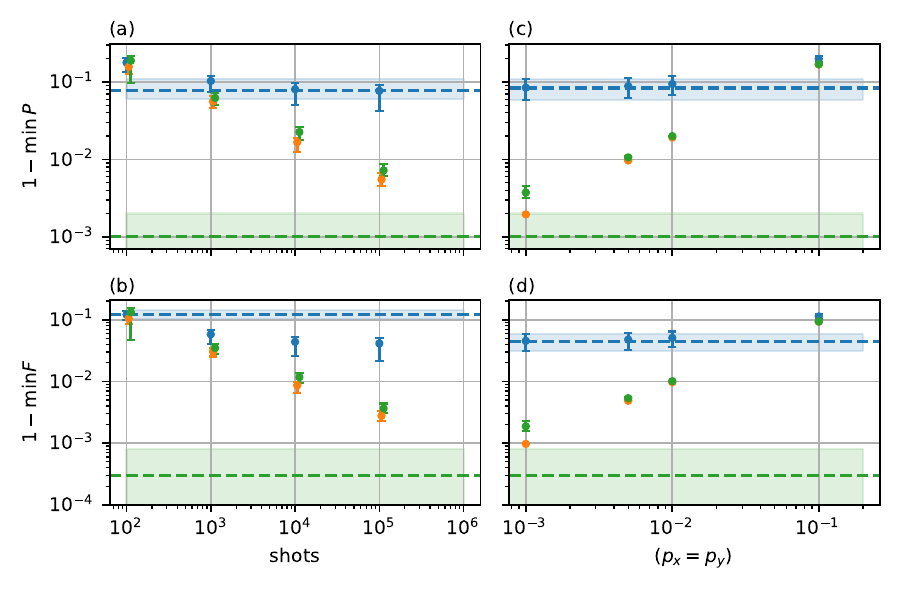}
	\caption{Performance limitations in the presence of (a, b) shot noise and (c, d) environmental noise. Calibration performance (green) is compared to two edge cases: a naive approach assuming zero error parameters (blue) and a scenario with perfect knowledge of error parameters (orange). Horizontal bands indicate the zero-noise limit. The values at the zero-noise, with perfect knowledge of the error parameters, are limited by the numerical precision of our reconstruction algorithm, given the finite number of iterations. Specifically, the green bands visualize the results from Section~III of the main text, and the blue band represents a reference case of noiseless tomographies without calibration. Points show the median maximal impurity (a,c) and infidelity (b,d) in the test ensemble across 50 Monte Carlo realizations; error bars represent the 0.158 and 0.842 quantiles. Slight horizontal shifts of the datasets are for visualization only.}
	\label{sfig:0}
\end{figure*}

To evaluate the method performance, we again use the following figures of merit: maximal impurity and maximal infidelity of the test ensemble. The discussion of infidelity is at the end of this section. As shown in Figure~\ref{sfig:0}(a,b), starting from roughly $10^3$ shots per projection, the proposed method already provides an advantage: the impurity decreases to $1 - P = (62^{ +12}_{-10 }) \times 10^{ -3 }$ and the infidelity to $1 - F = (35^{ +7}_{-6 }) \times 10^{ -3 }$. 

This is in contrast to the naive approach, which yields $1 - P = (10^{ +3}_{-2 }) \times 10^{ -2 }$ and $1 - F = (58^{ +17}_{-10 }) \times 10^{ -3 }$. For comparison, using perfect knowledge of the true error parameters leads to $1 - P = (56^{ +10}_{-10 }) \times 10^{ -3 }$ and $1 - F = (30^{ +5}_{-5 }) \times 10^{ -3 }$. The error bars represent the 0.158 and 0.842 quantiles, corresponding approximately to one standard deviation. The impurity and infidelity for the reference case of perfectly knowing the measurement operators are nonzero because we still simulate the test tomograms with a finite number of shots.

We now turn to the performance under environmental noise. In this model, a qubit undergoes a bit-flip with probability $p_x/2$ and a phase-flip with probability $p_z/2$. A mixing factor of $p_z = 1$ or $p_x = 1$ corresponds to full dephasing or full depolarization, respectively. This noise can be described using Kraus operators:
\[
\rho \rightarrow \sum\limits_{i=0}^{3}p_i K_i \rho K_i^{\dagger},
\]
where the parameters are defined as:
\[
\begin{aligned}
	p_0 &= (1 - p_x/2)(1 - p_z/2), & K_0 &= \mathbb{I}, \\
	p_1 &= (p_x/2)(1 - p_z/2), & K_1 &= \sigma_x, \\
	p_2 &= (1 - p_x/2)(p_z/2), & K_2 &= \sigma_z, \\
	p_3 &= (p_x p_z)/4, & K_3 &= \sigma_x \sigma_z.
\end{aligned}
\]

This type of noise breaks the assumption that all prepared states have the same purity and is, therefore, expected to limit the method’s performance. We reran the simulation from Section~III, modifying it so that all probe states are affected by this environmental noise. The mixing factors $p_x = p_z$ were gradually increased from 0 to 0.1 in logarithmic steps, and for each step, 50 Monte Carlo realizations were taken.

Figure~\ref{sfig:0}(c,d) shows the resulting degradation in performance. At $p_x = p_z = 0.1$, the method can no longer accurately estimate the error parameters, and its performance becomes comparable to that of the naive approach.

At a lower noise level of $p_x = p_z = 10^{-2}$, however, the method still yields significant improvement over the naive approach, from $1 - P = (9^{ +3}_{-2 }) \times 10^{ -2 }$ to $(200^{ +5}_{-3 }) \times 10^{ -4 }$ and from $1 - F = (5^{ +2}_{-1 }) \times 10^{ -2 }$ to $(101^{ +3}_{-2 }) \times 10^{ -4 }$, almost reaching the theoretical limit (i.e., reconstruction using true error parameters) $1 - P = 1.9 \times 10^{-2}$ and $1 - F = 9.7 \times 10^{-3}$. 

Note that no confidence interval is given for reconstructions assuming perfect knowledge of the measurement operators, as all the Monte Carlo tomograms are reconstructed perfectly, regardless of the true error parameters. The purity and fidelity limitation in this case are determined by the dephasing and depolarization affecting the prepared states, which remain the same for all Monte-Carlo samples.

To assess fidelity without resorting to the optimization described in Section~III of the main text, we use the Kabsch-Umeyama algorithm~\cite{Kabsch1976, Umeyama1991} to align the dominant Bloch eigenvectors of the reconstructed ensemble with those of the ideal pure reference test states. Since the reference states are pure, the fidelity simplifies to:
\[
F = \kappa f + (1 - \kappa)(1 - f),
\]
where $f = \frac{1 + \vec{u} \cdot \vec{\kappa}}{2}$, and $\vec{\kappa}$ is the Bloch vector representation of the dominant eigenstate, with eigenvalue $\kappa$, corresponding to the test state. The vector $\vec{\kappa}$ is the corresponding eigenvector, while $\vec{u}$ is the Bloch vector of the reference state. This expression gives the fidelity that would be achieved after optimally aligning the reconstructed test ensemble to the ideal one. To describe the calibration performance, we always report the highest infidelity of the test ensemble.

Although we have investigated only two representative noise models here, the procedure outlined in this section can be readily applied to assess the calibration performance under any noise model of interest.

Given a specific noise model, one could further improve the tolerance to such noise by tailoring the optimization cost function. For example, in the case of pure dephasing, the purity decreases as we approach the Bloch equator. States that share the same latitude would theoretically have equal purity. Reconstruction artifacts, however, would break this uniformity. Therefore, instead of measuring purity modulation over the entire Bloch sphere, one could quantify the purity modulation within equi-latitude bands and use that as the cost function.

\section{\label{sec:scaling}Applicability in a multi-qubit scenario}
The calibration of individual qubits is also advantageous in multi-qubit scenarios. Starting with a worst-case, let state $\ket{\psi_{i}}$ be the most sensitive state at the $i$-th qubit. Then, product state $\bigotimes\limits_{i}\ket{\psi_{i}}$ is the $n$-qubit state with the greatest manifestation of the artifact, with $P_{\min, n} = \prod\limits_{i=1}^{n} P_{\min,i}$, where $P_{\min,i}$ is the corresponding minimal purity observed for the single qubit. However, the important question arises whether the calibration provides better overall results in a multi-qubit scenario.

\begin{figure}[ht]
	\includegraphics[scale=1]{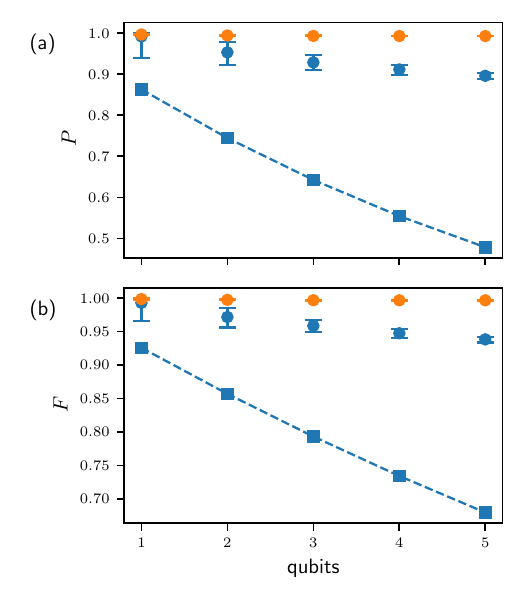}
	\caption{Scaling of reconstruction artifacts in a multi-qubit scenario. Data show median and 0.158 and 0.842 quantiles of reconstructed purity and fidelity of 500 Haar-random states. The reference for non-calibrated reconstructions is shown as blue dots. Applying the knowledge of true $\vec{\delta}$ improves the reconstruction quality in all tested multi-qubit cases, as shown using orange dots. For comparison, the blue squares illustrate the most sensitive state without calibration. In all tested cases with calibration, the purity and fidelity virtually reach one, and the achieved value is limited only by the number of iterations in the reconstruction algorithm used.}
	\label{sfig:multi-qubit-scaling}
\end{figure}

To assess the utility of the calibration in the multi-qubit setting, we performed the following numerical simulation. For a given number of qubits, one to five, we drew 500 Haar-random pure states and simulated their tomograms with a certain error vector $\vec{\delta} \neq 0$. We set the $\vec{\delta}$ to be the same for all qubits for ease of comparison. We then simulated the tomographical measurements, assuming no noise. We have analyzed the purity and fidelity of the resulting reconstructed density operators. We did the reconstructions twice, first with choice $\vec{\delta}' = \vec{0}$, i.e. assuming no systematic error. This served as a reference case. Then, we performed the calibration of each individual qubit, meaning we have gained knowledge of the true error vector, and we chose $\vec{\delta}' = \vec{\delta}.$ The improvement in purities and fidelities, shown in  Fig.~\ref{sfig:multi-qubit-scaling}, clearly indicates that individual calibration brings an advantage in multi-qubit scenarios as well, even though the overall artifacts are not as pronounced as in the worst case. In all tested cases, the calibration improved the purity and fidelity to almost one. The remaining infidelity and impurity are related to the finite iteration number in the maximum-likelihood reconstruction algorithm that we use and could be further reduced by allowing more iteration steps and setting smaller stopping criteria.


\begin{figure}[ht]
	\includegraphics[width=0.99\linewidth]{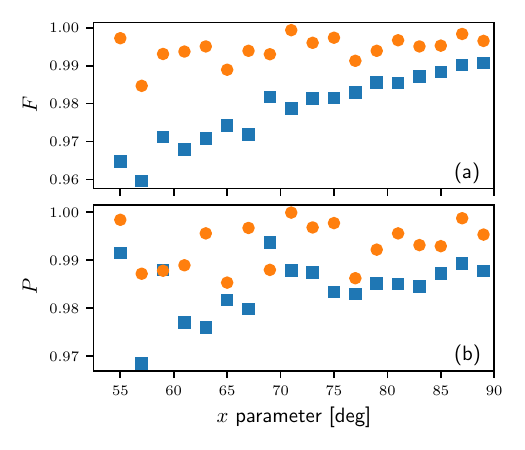}
	\caption{Two-qubit state purity (a) and fidelity (b) before (blue) and after (orange) calibration of individual qubits. Parameter $x$ determines which state we prepare and is very specific to experiment~\cite{Horova2022}. In short it determines the population of the first qubit, while the population of the second qubit is coupled to this parameter.}
	\label{fig:sfig-battery-experiment}
\end{figure}

Furthermore, we show how we have utilized the calibration method in the two-qubit photonic experiment~\cite{Horova2022} aiming to experimentally investigate mutual coherence on 2D, 3D, and 4D Hilbert subspaces. The qubits were encoded into polarizations of single photons emerging from the SPDC process. The calibration of individual qubits was utilized to extract the true effective retardances of the wave plates used in the experiment. Then we took this information into account while preparing and projecting the state, and it resulted in higher fidelity of the prepared two-qubit states. The results in Fig.~\ref{fig:sfig-battery-experiment} show that the presented method remains effective in the case of two-qubit experiments with single photons.

The utility of the calibration reaches beyond the reconstruction quality per se. The reduced deviation of the actual and the desired intermediate states in a quantum information protocol could result in better overall quality of the protocol. 
Even if the compensated error is small, it can be important when used multiple times. Here we present a minimal single-qubit experiment showing this effect. The qubit is initially prepared in state $|+\rangle$, evolved using some unitary $U$, in this case 
$$
U = e^{i\beta\sigma_x} e^{i\beta\sigma_y} e^{-i\beta\sigma_x} e^{-i\beta\sigma_y},
$$
and subsequently tomographically analyzed. The next iteration begins by preparing the dominant eigenket of the reconstructed density matrix and the cycle continues. The experimental setup is the same as in the main text, up to an extra stack of quarter-, half-, and quarter-wave plates in the central part. The central stack implements the evolution step $U$, which we have characterized using quantum process tomography. 

\begin{figure}[ht]
	\includegraphics[width=0.75\linewidth]{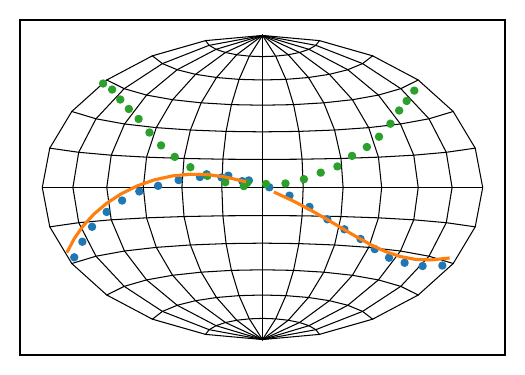}
	\caption{Experimental simulation of quantum dynamics. When using calibration, the experimental state indicated as blue dots, starting from state $|+\rangle$, closely follows the theoretically predicted trajectory, depicted as the orange line. Repeating the experiment without calibration results in a green trajectory, which diverges from the theoretical after a few iterations. The qubit is initialized in state $|+\rangle$ and evolves with a step $U = e^{i\beta\sigma_x} e^{i\beta\sigma_y} e^{-i\beta\sigma_x} e^{-i\beta\sigma_y}$, where $\beta = \pi/12$, and it gradually moves westward on the Bloch sphere.}
	\label{fig:sfig-evolution}
\end{figure}

We performed such an experimentally simulated evolution with and without calibration, and the resulting trajectories are compared in Fig.~~\ref{fig:sfig-evolution}. It indicates that with calibration, the state closely follows the theoretically predicted trajectory, while without calibration, the state diverges from that trajectory after a few iterations.

\section{\label{sec:s1o}Artifact mitigation of multiplicative errors}

Here, we explore the sampling strategies in the case of multiplicative errors. Since the artifact is state-dependent, the probe sampling strategy is important for minimizing the reconstruction artifacts. 

As in the main text, all states are measured in projections onto eigenstates of Pauli operators. The projections are implemented by rotating the measured state using the following unitary operation $\hat{R}_{y}(\theta_j)\hat{R}_{z}(\phi_j)$ and subsequent projection of the resulting state onto state $|0\rangle$. The rotation angles $\theta_j, \phi_j$ determine the selected projection and are listed in Tab~\ref{tab1}.

\begin{figure*}[ht]
	\centering
	\includegraphics[width=0.95\linewidth]{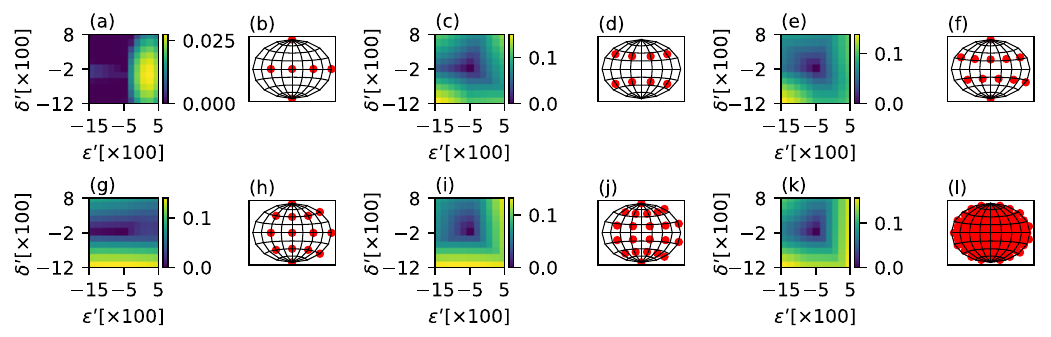}
	\caption{Dependence of simulated artifact severity, represented as purity modulation amplitude depicted as 2D color maps in panels (a, c, \dots, k) on the selection of probe states set, depicted as points on projected Bloch sphere in panels (b, d, \dots, l). The figure shows the case for 6 (a,b), 8 (c,d), 12 (e,f), 14 (g,h), 22 (i,j), and 108 (k,l) probe states. The purity modulation amplitude is always plotted around the ground truth $\delta = 0.02$, $\delta = -0.04$.}
	\label{sfig:1}
\end{figure*}

We test a few sampling schemes and plot the cost function near the ground truth ($\delta = 0.02$, $\epsilon = -0.04$) for each. The cost function is the peak-to-valley value of reconstructed probe state purities, $\Delta P = P_{\max} - P_{\min}$. The probe sets and cost functions are depicted in Figure~\ref{sfig:1}. As probe states set, we selected eigenstates of Pauli operators (panels a,b),  eight states in (panels c, d), icosahedron (panels e, f), and then a uniform sampling of Bloch sphere angular coordinates ($\theta, \phi$) with 14 states (panels g, h) and 22 states (panels i, j) and finally NASA HEALPIX~\cite{gorski2005, hardin2016} with 108 probe states (panels k, l). 

\begin{table}[b]
\begin{tabular}{|c|c|c|c|c|c|c|}
	\hline
	$j$ & 1 & 2 & 3 & 4 & 5 & 6 \\
	\hline
	$\theta_j$ ($y$-axis) & 0 & $\pi$ & $\pi/2$ & $\pi/2$ & $\pi/2$ & $\pi/2$ \\
	$\phi_j$ ($z$-axis) & 0 & 0 & $\pi$ & 0 & $\pi/2$ & $3\pi/2$ \\
	\hline
\end{tabular}	
	\caption{Rotation angles in radians for Pauli tomography. The analyzed state is first rotated by $\phi_j$ around the $z$-axis and $\theta_j$ around the $y$-axis and subsequently projected to state $|0\rangle$.}
	\label{tab1}
\end{table}

\begin{figure*}[pt]
	\includegraphics[width=0.6\linewidth]{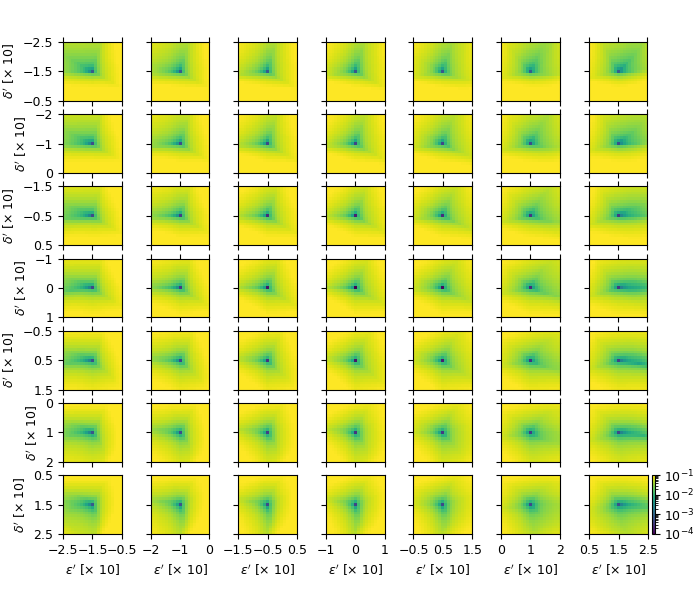}
	\caption{The panels show color-coded purity modulation as a function of assumed error parameters ($\delta', \epsilon'$) observed in the set of eight reconstructed probe states. The purity modulation is sampled near the true value ($\delta, \epsilon$) for each simulated ground truth. This true value is always plotted in the center of the panel. All panels share the color bar. To visually emphasize the presence of a single minimum, we use logarithmic normalization of color coding.}
	\label{sfig:2}
\end{figure*}

\begin{figure*}[pb]
	\includegraphics[width=0.6\linewidth]{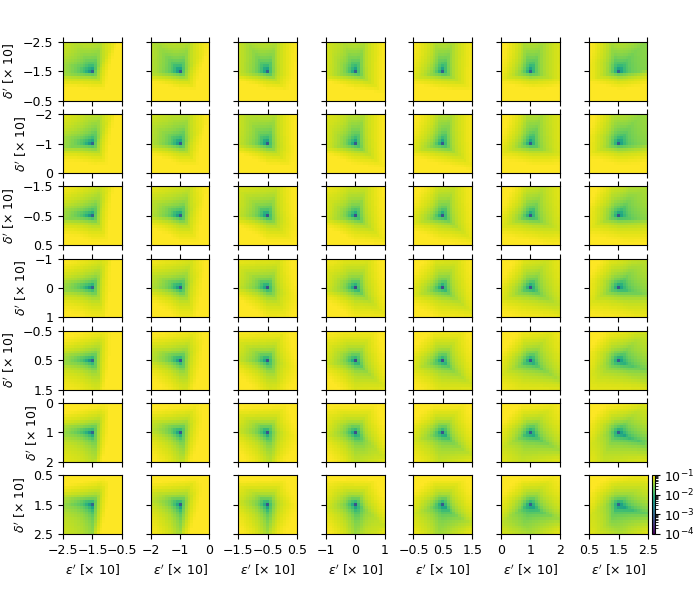}
	\caption{12 probe states. The figure is arranged as Fig.~\ref{sfig:2}.}
	\label{sfig:3}
\end{figure*}

\begin{figure*}[p]
	\includegraphics[width=0.6\linewidth]{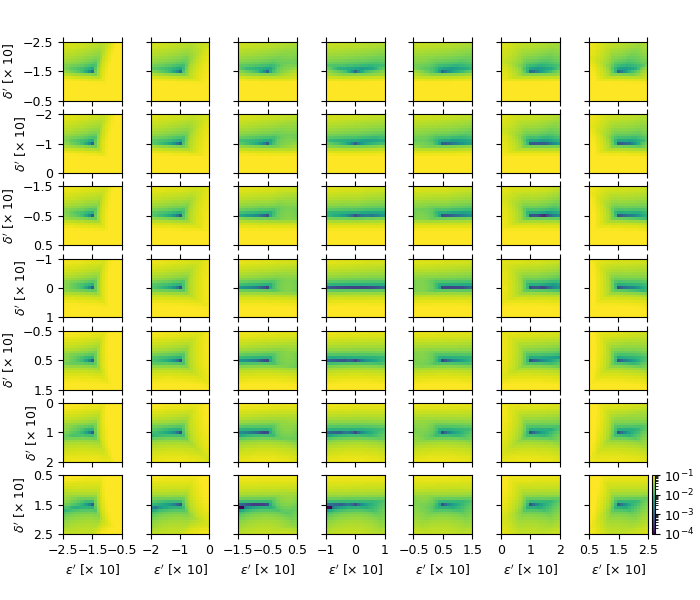}
	\caption{14 probe states. The figure is arranged as Fig.~\ref{sfig:2}.}
	\label{sfig:4}
\end{figure*}

\begin{figure*}[p]
	\includegraphics[width=0.6\linewidth]{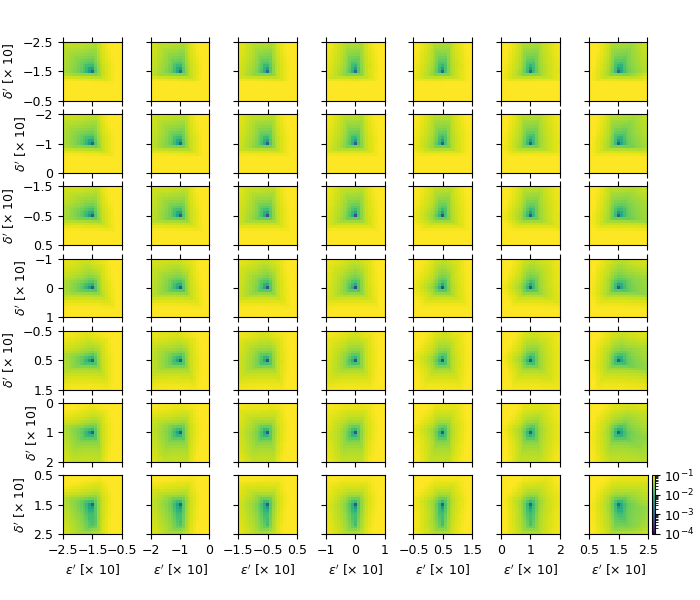}
	\caption{22 probe states. The figure is arranged as Fig.~\ref{sfig:2}.}
	\label{sfig:5}
\end{figure*}

\begin{figure*}
	\includegraphics[width=0.6\linewidth]{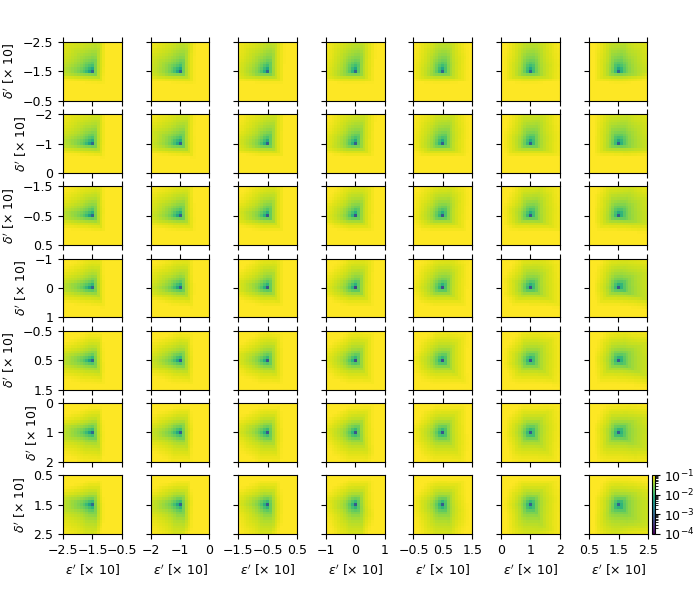}
	\caption{108 probe states. The figure is arranged as Fig.~\ref{sfig:2}.}
	\label{sfig:6}
\end{figure*}

The sampling is sufficient when the cost function possesses a unique global minimum. We have run additional simulations, varied the ground truth, and observed the shape of the cost function. We tested each sampling scheme from Fig.~\ref{sfig:1}. We sampled the cost function near the varying ground truth values of $\delta$ and $\epsilon$ and plotted its value as color maps, depicted in Fig.~\ref{sfig:1}. The first suitable sampling scheme appears for eight probe states. Interestingly, for 14 probe states, there is a rift in the cost function landscape that causes ambiguity in the $\delta, \epsilon$ determination. The remaining sampling schemes exhibit a well-defined global minimum and therefore are suitable. The cost function landscapes in Fig.~\ref{sfig:1} are plotted for a single choice of $(\delta, \epsilon)$. To better characterize the sampling strategies, we repeated the same simulations for multiple ground truths. Figures~\ref{sfig:2}-\ref{sfig:6} show the simulated cost function for 8, 12, 17, 22, and 108 probe states. The ground truth is always plotted in the panel center. This additional characterization confirms the result -- 8 probe states are enough and there is indeed anomalous behavior for 14 probe states. For 22 states, we sometimes see a local minimum, such as for ground truth parameters $\delta = 0.15$, $\epsilon = 0$. Due to this, the optimization has to be done carefully, ideally with several starting points, to find the global minimum of the cost function. In the case of 108 probe states, the cost function always had a single minimum.

The most interesting outcome of this analysis is that the success of the sampling strategy is not determined by the sheer number of probe states but rather by their suitable placement on the Bloch sphere. Generally, adding more probe states makes the error parameters determination more robust because we increase the obtained information, however, at the cost of longer measurement time. The exact cost function shape, sampling strategies, and shapes of the purity modulation depend on the imperfection type, its parameterization, and how we switch the measurement operators in the tomography. 

\begin{figure*}[ht]
	\centering
	\includegraphics[scale=1]{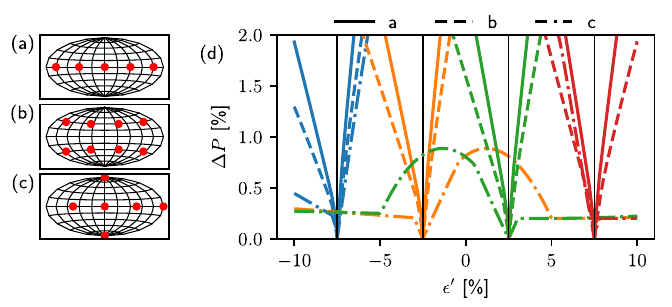}
	\caption{Influence of the probe set choice on the cost function. The tested probe sets are depicted in panels (a), (b), and (c), and their corresponding cost functions are plotted using solid, dashed, and dash-dotted lines, respectively. The cost function was evaluated for four distinct true values of $\epsilon$, denoted by black vertical lines. Colors are used to distinguish between the various tested true values of $\epsilon$. The six-state probe set is not suitable for calibration.}
	\label{sfig:7}
\end{figure*}

We further illustrate this behavior using a simple example of phase over-rotation. In this case, $\delta = 0$, while $\epsilon \neq 0$. The reconstruction artifacts appear most prominently near the equator of the Bloch sphere. One might argue that the optimal strategy is to place the probe states uniformly along the equator, as depicted in Fig.~\ref{sfig:7}(a). We ran simulations using five probe states placed uniformly on the Bloch equator and plotted the purity modulation as a function of the assumed $\epsilon$ for four distinct values of the true error factor: $\epsilon = \pm 2.5\%$ and $\epsilon = \pm 7.5\%$. The corresponding cost function is shown in Fig.~\ref{sfig:7}(d) using solid lines.

As a reference, we also plotted the purity modulation for two additional probe sets: eight quasi-uniformly placed states (Fig.~\ref{sfig:7}(b), dashed lines) and the six Pauli eigenstates (Fig.~\ref{sfig:7}(c), dash-dotted lines). Indeed, the equatorial probe state set yields a clearer minimum than using six states. In this simple example, we demonstrate that, given the error model, it is  possible to optimize the probe set to achieve improved calibration performance and achieve conclusive results with a smaller probe state set.

\begin{table}[t]
	\begin{tabular}{|c|c|c|c|c|c|c|}
		\hline
		state & $|H\rangle$ & $|V\rangle$ & $\frac{|H\rangle + |V\rangle}{\sqrt{2}}$ & $\frac{|H\rangle - |V\rangle}{\sqrt{2}}$ & $\frac{|H\rangle + i|V\rangle}{\sqrt{2}}$ & $\frac{|H\rangle - i|V\rangle}{\sqrt{2}}$ \\	
		$j$ & 1 & 2 & 3 & 4 & 5 & 6 \\
		\hline
		$x_j$ (HWP) & 0 & $\pi/4$ & $\pi/8$ & $-\pi/8$ & $-\pi/8$ & $\pi/8$ \\
		$y_j$ (QWP) & 0 & 0 & 0 & 0 & $-\pi/4$ & $\pi/4$ \\
		$x'_j$ (HWP) & 0 & $\pi/4$ & $\pi/8$ & $-\pi/8$ & $\pi/8$ & $-\pi/8$ \\
		$y'_j$ (QWP) & 0 & 0 & 0 & 0 & $\pi/4$ & $-\pi/4$ \\
		\hline
	\end{tabular}
\label{tab2}
\caption{Waveplate angular position for Pauli tomography.
    The waveplates are placed as follows in the direction of the light  propagation: quarter-wave plate and half-wave plate in the preparation,  and half-wave plate and quarter-wave plate in the measurement stage. Here, the primed symbols correspond to the state preparation used in the reversed version of the method.}
\end{table}

\section{\label{sec:s4}Rotating waveplate polarimetry}

\begin{figure*}[p]
	\centering
	\includegraphics[scale=1]{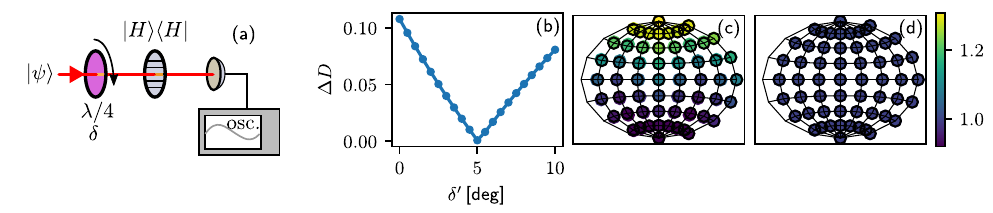}
	\caption{Mitigation of artifact in rotating-waveplate polarimetry. (a) Measurement apparatus. A quarter-wave plate (purple) with retardance deviation $\delta = 5$ deg is rotating in front of the linear polarizer. The photodiode signal is recorded and analyzed. (b) The minimal $\Delta D$ lies at the true retardance deviation $\delta$. (c) Initial mismatch results in a state-dependent purity artifact. We plot the reconstructed probe states on the projected Bloch sphere and color-code their reconstructed Stokes vector magnitude $D = \sqrt{S^2_x + S^2_y + S^2_z}/S_0$. (d) When we considered the true retardance, the artifact vanished. Black circles depict the ideal testing states.}
	\label{sfig:rotqwp}
\end{figure*}

Consider the classical rotating quarter-wave plate polarimeter~\cite{Goldstein}, depicted in Fig.~\ref{sfig:rotqwp}(a), with the quarter-wave plate having the true retardance deviated by $\delta = 5\deg$ from its nominal value. Moreover, consider pure probe states. The waveplate rotates continuously and effectively modulates the optical signal detected after the polarizer. Instead of likelihood maximization, we now use classical Fourier analysis to reconstruct the polarization state. We chose the Fourier analysis to show that our method is not limited to maximal likelihood reconstruction. We used eight probe states to detect the reconstruction artifact, and instead of purity modulation, we equivalently measured the severity with a modulation amplitude of the degree of polarization. 

First, we treat the data with the ideal retardance ($\delta' = 0$) assumption. We see that the states from the ensemble suffer from artifacts. Here, we look at the intensity-normalized Stokes vector length $D = \frac{\sqrt{\sum\limits_{i=x,y,z}S_i}}{S_0}$, which is proportional to purity $P = \frac{1}{2}\left(1 + D^2\right)$. Near state $\ket{H}$, at the north pole of the Bloch sphere, the purity even exceeds 1 due to the reconstruction method. The modulation of $D$ is depicted in Fig.~\ref{sfig:rotqwp}(c). This behavior is clearly a reconstruction artifact. Exceeding the unit purity is unsurprising because Fourier analysis does not guarantee physically sound results. We now vary $\delta'$ assumed in reconstruction and observe the peak-to-peak value $\Delta D$ in the probe set. The dependence is depicted in Fig.~\ref{sfig:rotqwp}(b). We see that the minimum lies at the true $\delta$. In panel (d), we show that the artifact disappears when the retardance deviation is taken into account.

\FloatBarrier
\bibliography{sreference}